\documentclass{l4dc2024}
\usepackage{algorithmic}
\usepackage{algorithm}
\usepackage{mathtools}
\title[FRIDAY]{FRIDAY: Real-time Learning DNN-based Stable LQR controller for Nonlinear Systems under Uncertain Disturbances}
\usepackage{times}

\author{%
 \Name{Takahito Fujimori} \Email{TakahitoFujimoriControls@gmail.com}\\
 \addr School of Biology, Osaka University, Machikaneyamacho 1-1, Toyonaka, Japan
}

\begin{document}

\maketitle

\begin{abstract}%
  Linear Quadratic Regulator (LQR) is often combined with feedback linearization (FBL) for nonlinear systems that have the nonlinearity additive to the input. Conventional approaches estimate and cancel the nonlinearity based on first principle or data-driven methods such as Gaussian Processes (GPs). 
  However, the former needs an elaborate modeling process, and the latter provides a fixed learned model, which may be suffering when the model dynamics are changing. 
  In this letter, we take a Deep Neural Network (DNN) using real-time-updated dataset to approximate the unknown nonlinearity while the controller is running. 
  Spectrally normalizing the weights in each time-step, we stably incorporate the DNN prediction to an LQR controller and compensate for the nonlinear term. 
  Leveraging the property of the bounded Lipschitz constant of the DNN, we provide theoretical analysis and locally exponential stability of the proposed controller. 
  Simulation results show that our controller significantly outperforms Baseline controllers in trajectory tracking cases. 
\end{abstract}


\section{Introduction} 
\indent Although Linear Quadratic Regulator (LQR) is an intuitively simple controller for linear systems and one of the successful theoretical subjects in modern control theory \citep{LQR}, 
many safety-critical systems such as self-driving vehicles, unmanned aerial vehicles (UAVs) and mobile manipulators exhibit additive nonlinear dynamics to the control input \citep{Path}. 
For such systems, LQR is often combined with feedback linearization (FBL) to cancel out the nonlinearity, and various identification methods have been proposed.\\ 
\indent A classical approach is first principle based \citep{1stPri} or adaptive \citep{MRAC} approximations. However, parametric FBL-LQR controllers often suffer from slow response and delayed feedback if the model mismatches with the true dynamics. 
Another approach utilizes data-driven estimation, such as Gaussian Processes (GPs). GPs learn the nonlinear term and present the quantified prediction mismatch, enabling GPs-FBL-LQR controllers to stably linearize the system \citep{GPsFilter, GPsFlatness}. 
However, GPs often keep a fixed dataset and do not update the model while the controller is running. Fixed models may suffer from situations where the model dynamics are changing due to unexpected disturbances, or the model needs to be updated because of biased training data. 
The compensating strategy is adding data to the dataset in real-time \citep{SecTri} or updating models based on its current reliability \citep{EveTri}, but these methods heavily rely on data efficiency and need a possibly expensive online optimization process, compromising fast convergence to the reference states. 
This is because the data efficiency of GPs is very sensitive to the choice of kernel/covariance function, and the runtime complexity is inherently $O(n^3)$. 
For these reasons, we take Deep Neural Networks (DNNs) approach to compensate for the unknown disturbances in real-time.\\  
\indent \textit{Related works:} 
\citep{impromptu, ActSus} incorporate a learned DNN into a controller and directly cancel out the uncertainty by its prediction. However, these DNNs are not theoretically analyzed and thus possibly generate unpredictable outputs.
To guarantee stability, \citep{NL, Ocean} apply Spectral Normalization (SN) to their DNN weights to constrain the Lipschitz constants. Though they stably compensate for complex fluid dynamics, they have fixed training datasets, resulting in the dependency of the performance on the DNN generalizations.
Concurrent with real-time execution, \citep{iter1,iter2} collect data and iteratively train their DNNs with Stochastic Gradient Descent (SGD) to apply batch-like updates. Still, in such multi-time scale controllers, when to collect the data and when to update the weights are open questions.
Instead of SGD, adaptive DNN weights update laws based on Lyapunov stability analysis have been developed to continuously adjust the weights. Though these architectures are well-established, they only apply to NNs with a single hidden layer \citep{single} or only update the output layer weights  \citep{outlayer}.
For full layer weights update, \citep{modular,derived} develop modular adaptive law, but in arbitrary width and depth DNNs, simultaneously updating weights online under adaptive laws may be computationally intractable or undesired.\\ 
\indent \textit{Contributions:}
Exploiting the technique to stabilize DNN-based FBL signal \citep{NL}, which is computationally light and easy to implement, we continuously update all the weights with simple SGD, not with adaptive laws. 
Specifically, we apply SN to the weights before executing our DNN-based control input in each time-step such that it be a contraction mapping, converging to its unique point. By doing so, we can stably optimize the weights with SGD while controlling the system. 
The proposed controller is named FRIDAY, short for Fast ResIdual Dynamics AnalYsis. Leveraging the property of the bounded Lipschitz constant, FRIDAY is proved to be locally exponentially stable under bounded learning error.
Simulation results show FRIDAY achieves almost double trajectory tracking accuracy of an adaptive baseline controller and ten times that of an LQR, while learning the map of uncertain disturbances. 
To the best of our knowledge, this is the first guaranteed framework that constantly collects data and updates the full layer weights with SGD to cancel out unknown dynamics.

\section{Problem Statement: Nonlinear Systems under Uncertain Disturbances}
\indent Given the states of a system as state vector $\mathbf{x} \in \mathbb{R}^{\text{n}} $, input vector $\mathbf{u} \in \mathbb{R}^{\text{m}} $, 
we consider a control-affine nonlinear dynamic system
\begin{equation}
\label{Non}
\dot{\mathbf{x}}=f(\mathbf{x})+g(\mathbf{x})\mathbf{u}.
\end{equation}
A wide range of dynamical systems such as quadrotor and car-like vehicles can be separated into a linear dynamics component and additive nonlinearity \citep{GPsFilter,GPsFlatness}. Thus, we divide the nonlinear system into a Linear Time-Invariant (LTI) system and a nonlinear term as follows \citep{Path}:
\begin{equation}
\label{Prob}
\dot{\mathbf{x}}= A\mathbf{x}+B(\mathbf{u}+\mathbf{R}(\mathbf{x}, \mathbf{u})),
\end{equation}

\noindent where $A,B$ are time-invariant matrices with the pair $(A,B)$ being controllable, and $\mathbf{R}(\mathbf{x}, \mathbf{u})$  accounts for unknown nonlinear dynamics including model uncertainties, named residual dynamics. It is noted that the analysis in this letter
is restricted to this form.

\indent \textit{Problem Statement}: We aim to make a copy of the residual dynamics $\hat{\mathbf{R}}(\mathbf{x}, \mathbf{u})$ in real-time to cancel out the nonlinear term by control input $\mathbf{u}=\mathbf{u}^{\text{LQR}} - \hat{\mathbf{R}}$ such that the LQR controller becomes able to operate the linearized system. 
We use a real-time learning DNN directly as the cancellation term, which means we incorporate DNN predictions into a controller while the DNN is learning. 
To stabilize the closed-loop, we exploit contraction mapping technique proposed by \citep{NL} using Spectral Normalization, which constrains the Lipschitz constant of DNNs. 
Leveraging the property of the bounded Lipschitz constant, we guarantee the contoller stability.

\begin{figure}
  \centering
  \includegraphics[width=0.5\linewidth]{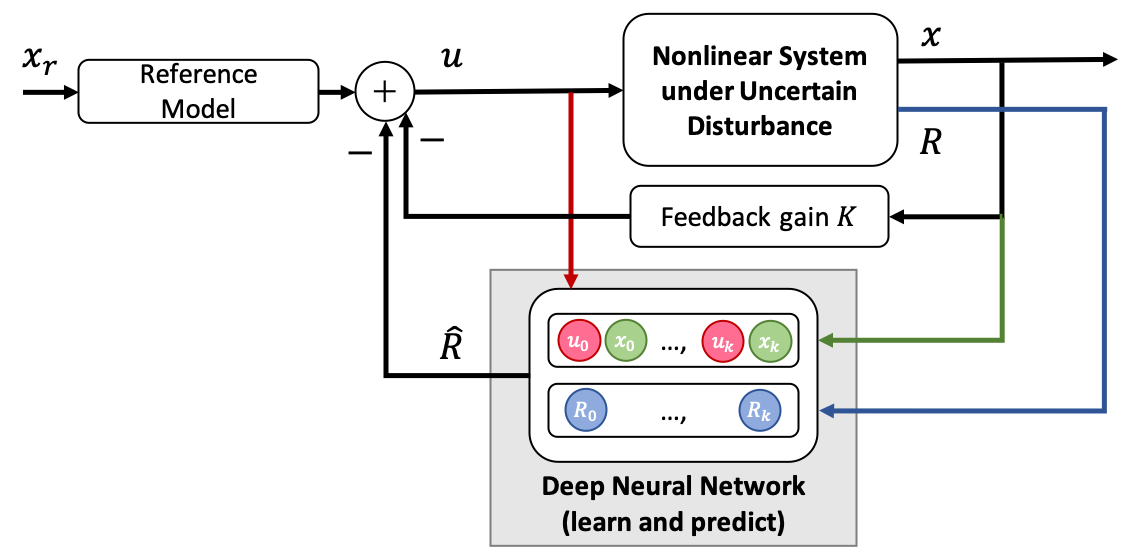}
  \caption{Our proposed architecture using real-time learning DNN has three key components: (1) Updating Dataset in Real-time: the current observed values of states, control inputs and residual dynamics are added to the training dataset in each time-step, (2) One-time Learning: using the dataset, the DNN is optimized once an iteration with Stochastic Gradient Descent, (3) Execution after Spectral Normalization: the DNN estimation value cancels the nonlinearity after its Lipschitz constant is bounded.}
  \label{DIA}
\end{figure}

\section{FRIDAY Controller Design}
\indent We first show the overall design of our controller, composed of a conventional LQR and the DNN-based feedforward cancellation $\hat{ \mathbf{R}}(\mathbf{x}, \mathbf{u}) $. The DNN learns $\mathbf{R}$ using observed values while its prediction directly compensates for the unknown dynamics such that the LQR controls the system (see Fig. \ref{DIA})
\begin{equation}
\label{FRIDAY}
\mathbf{u}= -K(\mathbf{x}-\mathbf{x}_r)+\mathbf{u}_{r} -\hat{ \mathbf{R}}(\mathbf{x}, \mathbf{u}),
\end{equation}
\noindent where the feedback gain $K = R^{-1}B^\top P$ derives from the solution of the Algebraic Riccati Equation (ARE) $A^\top P + PA -PBR^{-1}B^\top P + Q = 0, Q, R >0$. 
$\mathbf{x}_r$ and $\mathbf{u}_{r}$ are reference signals satisfying $\dot{\mathbf{x}}_r= A\mathbf{x}_r+ B\mathbf{u}_{r}$ \citep{LQR}. Substituting (\ref{FRIDAY}) into (\ref{Prob}), the closed-loop dynamics would be the following equation with approximation error $\boldsymbol{\epsilon}=  \mathbf{R}-\hat{ \mathbf{R}}$,
\begin{equation}
\label{FRItoSys}
\dot{\mathbf{x}}-\dot{\mathbf{x}}_r= (A-BK)(\mathbf{x}-\mathbf{x}_r) +B\boldsymbol{\epsilon}.
\end{equation}
\noindent By defining $\mathbf{z}=\mathbf{x}-\mathbf{x}_r$, $A_{\text{cl}}=A-BK$, the system dynamics would simply be
\begin{equation}
\label{FRItoSys2}
\dot{\mathbf{z}}= A_{\text{cl}}\mathbf{z} +B\boldsymbol{\epsilon}.
\end{equation}
\noindent As long as $\| B\boldsymbol{\epsilon} \| $ is bounded, $\mathbf{x}(t) \to \mathbf{x}_r(t)$ locally and exponentially with the bounded error \citep{slotinebook, ambulance}.

\section{DNN Learning Residual Dynamics}
\indent We demonstrate how the DNN is trained and predicts the residual dynamics while the controller is running. 
We use a DNN with Rectified Linear Units (ReLU) activation function. ReLU DNNs have been shown to converge faster, have fewer vanishing gradient problems, and be easier to optimize than the other activation functions such as \textit{sigmoid} and \textit{tanh} \citep{ReLU1} 
\begin{equation}
  \label{WhatDNN}
  \hat{\mathbf{R}}(D_{X},\boldsymbol{\theta})=W^{L}a(W^{L-1}(a(W^{L-2}(\ldots a(W^{1}D_{X})\ldots)))),
\end{equation}
\noindent where $D_{X}=\{\mathbf{x}, \mathbf{u} \}$ consists of the observed state and control input, $\boldsymbol{\theta}=W^{1},\cdots,W^{L}$ stands for $L$-th layers of weight matrices, and $a(\cdot)=\text{max}(\cdot,0)$ is layer-wise ReLU.

\subsection{Real-time Learning}
\indent Real-time learning means that the following optimization is conducted in each time-step
\begin{equation}
\begin{aligned}
\label{Opti}
\mathcal{D}_{X}&:= \Big\{  D_{X_{1}}, \cdots D_{X_{k-1}}\leftarrow + D_{X_{k}}=\{\mathbf{x}_{k}, \mathbf{u}_{k} \}  \Big\},\\
\mathcal{D}_{Y}&:= \Big\{  D_{Y_{1}}, \cdots D_{Y_{k-1}}\leftarrow + D_{Y_{k}}= \tilde{\mathbf{R}}_{k}  \Big\},\\
&\min_{\boldsymbol{\theta}} \quad  \frac{1}{\text{n}(N)}\sum_{n \in N}\| D_{Y_{n}}-  \hat{ \mathbf{R}}(D_{X_{n}},\boldsymbol{\theta})     \|^2   ,\\
\end{aligned}
\end{equation}
\noindent where $\tilde{\mathbf{R}}_{k}$ is the observed residual dynamics value and $N$ denotes a mini-batch set with its size being $\text{n}(N)$. 
In other words, the DNN is trainded using the each-iteration-updated dataset  $\mathcal{D}_{X}$, $\mathcal{D}_{Y}$. 
Once (\ref{Opti}) is done, the estimator $\hat{\mathbf{R}}$ is used for the cancellation at the next time-step (see Algorithm. \ref{alg1} line: 10 to 16, and 8).\\ 
\indent However, it is not preferable to integrate the estimation of such developping DNNs with a feedback controller because the output is unpredictable and can be unstable. To address this instability, 
we apply Spectral Normalization to our mapping.
\color{black}
\subsection{Spectral Normalization}
\indent Spectral Normalization (SN) normalizes the Lipschitz constant of the objective function. Lipschitz constant is defined as the smallest value such that
\begin{equation}
\label{WhatLip}
\forall \mathbf{x},\acute{\mathbf{x}} : \| f(\mathbf{x})-f(\acute{\mathbf{x}})  \|_{2} /  \| \mathbf{x}-\acute{\mathbf{x}} \|_{2} \leq \| f \|_{\text{Lip}}.
\end{equation}
\noindent Since the Lipschitz constant of the linear mapping $W\mathbf{x}$ is the spectral norm of the weight matrix $\sigma(W)$ (the maximum
singular value), and that of ReLU $\|a(\cdot)\|_{\text{Lip}}=1$, the Lipschitz constant of ReLU DNNs is naturally upper bounded by the product of all the spectral norms,
\begin{equation}
\label{InhLip}
\| f\|_{\text{Lip}} \leq \| W^{L}\|_{\text{Lip}}\cdot  \| a \|_{\text{Lip}} \ldots \| W^{1} \|_{\text{Lip}}= \prod_{l=1}^{L} \sigma(W^{l}).
\end{equation}
\indent Leveraging this property, we can make the Lipschitz constant of the DNNs upper bounded by an intended value $\zeta$ by dividing each weight $W_{SN}=  W^{l}/\sigma(W^{l}) \cdot \zeta^{ \frac{1}{L} }$ (see Sec. 2.1 \citep{SN}, \textit{Lemma 3.1} \citep{NL}).

\subsection{Constrained Prediction}
According to \textit{Lemma 5.1} and \textit{Theorem 5.2} discussed later, $\zeta$-Lipschitz DNN-based control input converges. 
Hence, stably incorporating a real-time learning DNN into a controller, now becomes normalizing all the weights s.t. $\| \hat{ \mathbf{R}}\|_{\text{Lip}} \leq \zeta$ before executing the control input $\mathbf{u}$ (see Algorithm. \ref{alg1} line: 4 to 9). \color{black}
We optimize the DNN using Stochastic Gradient Descent with momentum (Momentum SGD).

\begin{figure}[!t]
 
  \begin{algorithm}[H]
      \caption{FRIDAY algorithm}
      \label{alg1}
      \begin{algorithmic}[1]
      \STATE Initialize the weights $\theta$ of the SN-DNN.
      
      \FOR{the entire duration}
      \STATE Obtain the current states $\mathbf{x}_{k}$
      \STATE Calculate the spectral norm of each weight matrix $\sigma(W)$ and divide
      \FOR{$l$}
      \STATE $W^{l} \leftarrow  W^{l}/\sigma(W^{l}) \cdot \zeta^{ \frac{1}{L} }$  
      \ENDFOR 
      \STATE Estimate the residual dynamics $\hat{ \mathbf{R}}_{k}=\hat{ \mathbf{R}}(\mathbf{x}_{k}, \mathbf{u}_{k-1} )$
      \STATE Execute the current control input $\mathbf{u}_{k}=-K\mathbf{x}_{k}-\hat{ \mathbf{R}}_{k}$
      \STATE \textbf{Begin the DNN training}:
      
      \STATE Add the current data to the training sets,\\ 
      \STATE $\mathcal{D}_{X}\leftarrow + \{\mathbf{x}_{k}, \mathbf{u}_{k} \}$\\ 
      \STATE $\mathcal{D}_{Y}\leftarrow + \tilde{\mathbf{R}}_{k} $
      \STATE Sample random mini-batch $N$ and update $\theta$ once to minimize\\ 
      \STATE $\frac{1}{\text{n}(N)}\underset{{n \in N}}\sum\| D_{Y_{n}}-  \hat{ \mathbf{R}}(D_{X_{n}},\boldsymbol{\theta})     \|^2  $

      \STATE \textbf{End the training}.
      \ENDFOR   
      \end{algorithmic}
      
  \end{algorithm}
  
\end{figure}

\section{Theoretical Guarantees}
\indent We analyze the closed-loop system to prove its stability and robustness. This analysis also provides insight into how to tune the hyperparameters of the DNN and the LQR controller to improve the performance.
Note that all norms $\|\cdot\|$ used later denote $L^2$ norm.

\subsection{Convergence of Control input}
\indent Using fixed-point iteration, we show the control input defined as (\ref{FRIDAY}) converges to a unique point when we fix all states.

\noindent \textbf{\textit{Lemma 5.1:}} Control input defined by the following mapping $\mathbf{u}_{k}=\mathcal{F}(\mathbf{u}_{k-1})_{k}$ converges to the unique solution satisfying $\mathbf{u}_{k}^{*}=\mathcal{F}(\mathbf{u}_{k}^{*})_{k}$ when $\| \hat{ \mathbf{R}}\|_{\text{Lip}} \leq \zeta$ and all states are fixed, 
\begin{equation}
  \label{ConM}
  \mathcal{F}(\mathbf{u})_{k}=-K(\mathbf{x}_{k}-\mathbf{x}_{r_{k}})+ \mathbf{u}_{r_{k}}-\hat{ \mathbf{R}}( \mathbf{x}_{k},\mathbf{u} ).
\end{equation}

\indent \textit{Proof:} Given all states fixed and $\forall \mathbf{u}_{1},\mathbf{u}_{2} \in \mathcal{U}$, with $\mathcal{U}$ being a compact set of feasible control input, the distance in $L^{2}$-space is

\begin{equation}
\begin{aligned}
\label{proofConM}
\| \mathcal{F}(\mathbf{u}_{1})_{k} - \mathcal{F}(\mathbf{u}_{2})_{k}\|&=
\| \hat{ \mathbf{R}}( \mathbf{x}_{k},\mathbf{u}_{1})-\hat{ \mathbf{R}}( \mathbf{x}_{k},\mathbf{u}_{2} )\|\leq L_{R}\| \mathbf{u}_{1}-\mathbf{u}_{2}  \|,
\end{aligned}
\end{equation}
\color{black}
\noindent where $L_{R}$ is the Lipschitz constant of the estimated dynamics $\hat{ \mathbf{R}}( \mathbf{x},\mathbf{u} )$. 
As long as we constrain $L_{R}$ s.t. $L_{R} \leqq \zeta \leqq 1 $ in every iteration, $\mathcal{F}(\cdot)$ is always a contraction mapping. 
Thus $k$-th input $\mathbf{u}_{k}$ mapped by $\mathcal{F}(\cdot)_{k}$ get close to its unique solution. $\quad\Box$ 

\subsection{Stability of FRIDAY Controller}
\indent To present the stability of FRIDAY controller, we make four assumptions.\\
\noindent \textbf{\textit{Assumption 1:}} There exists the maximum of the reference signals $x_{r_m} =\max_{t \geq t_0}\|\mathbf{x}_r(t)\|$, $u_{r_m} =\max_{t \geq t_0}\|\mathbf{u}_{r}(t)\|$\\  
\noindent \textbf{\textit{Assumption 2:}} The distance between one-step control input satisfies $\| \mathbf{u}_{k}-\mathbf{u}_{k-1}\| \leqq  \rho \| \mathbf{z} \|$ with a small positive constant $\rho$.\\
\indent The intuition behind this assumption is provided as follows: From (\ref{ConM}), the distance is inherently upper bounded,
\begin{equation}
  \label{Assum1}
  \Delta \mathbf{u}_{k} \leq \sigma(K)\Delta\mathbf{z}_{k} +\Delta\mathbf{u}_{r_{k}}+ L_{R}(  \Delta\mathbf{u}_{k-1} + \Delta\mathbf{x}_{k} ),
\end{equation}

\noindent with $\Delta(\cdot)_{k}= \|(\cdot)_{k}-(\cdot)_{k-1}\|$. Under the condition where the update rate of all states control and the reference signals are much faster than that of FRIDAY controller, 
in practice, we can safely neglect $\Delta\mathbf{z}_{k}, \Delta\mathbf{u}_{r_{k}}$ and  $\Delta\mathbf{x}_{k}$ in one update (see Theorem11.1 \citep{khalilbook}, e. g. the rate$>$100Hz \citep{NL,NF}), which leads to:
\begin{equation}
\label{neglect}
\Delta\mathbf{u}_{k} \leq L_{R}  (\Delta\mathbf{u}_{k-1} + c),
\end{equation}
where $c$ is a small constant of sum of the neglected variables, and $L_{R} <1$. So we can consider $\Delta\mathbf{u}_{k}$ has a small ultimate bound, and there exists a positive constant $\rho$ s.t. $\| \mathbf{u}_{k}-\mathbf{u}_{k-1}\| \leq \rho \|\mathbf{z} \| $.\\
\noindent \textbf{\textit{Assumption 3:}} Over the compact sets of feasible states and control inputs $\mathbf{x} \in \mathcal{X}, \mathbf{u} \in \mathcal{U} $, the residual dynamics $\mathbf{R}( \mathbf{x},\mathbf{u} )$ and its learning error $\boldsymbol{\epsilon}(\mathbf{x},\mathbf{u})=  \mathbf{R}( \mathbf{x},\mathbf{u} )-\hat{ \mathbf{R}}( \mathbf{x},\mathbf{u} )  $ have upper bound 
$R_{m}=\sup_{\mathbf{x} \in \mathcal{X}, \mathbf{u} \in \mathcal{U}} \|\mathbf{R}(\mathbf{x},\mathbf{u}) \|$, $\epsilon_{m}= \sup_{\mathbf{x} \in \mathcal{X}, \mathbf{u} \in \mathcal{U}} \|\boldsymbol{\epsilon}(\mathbf{x},\mathbf{u}) \|$. \\
\indent This assumption is strengthened by \citep{Assum3} showing SN-DNNs empirically generalize well to the set of unseen events having almost the same distribution as the training set.\\
\noindent \textbf{\textit{Assumption 4:}} The compact sets $\mathcal{X}=\bar{B}_{r_\mathcal{X}}(\mathbf{0}, r_\mathcal{X})$, $\mathcal{U}=\bar{B}_{r_\mathcal{U}}(\mathbf{0}, r_\mathcal{U})$ are closed balls of radiuses $r_\mathcal{X}, r_\mathcal{U}$, centered at the origin.



\indent Based on the assumptions, we prove the stability and robustness of the closed-loop system.\\
\noindent \textbf{\textit{Theorem 5.2:}} If $\mathbf{x}_0 \in \mathcal{X}$, $\mathbf{u}_0 \in \mathcal{U}$ and $r_\mathcal{X}, r_\mathcal{U}$ are larger than some constants, 
then the controller defined in (\ref{FRIDAY}) achieves $\mathbf{x}(t) \to \mathbf{x}_r(t)$ exponentially to an error ball $\bar{B}_{r}$ of radius $r$.\\
\indent \textit{Proof:} We select a Lyapunov function as $ \mathcal{V}(\mathbf{z})=\mathbf{z}^\top P \mathbf{z}$ where P is a positive definite matrix satisfying the ARE. Applying \textit{Assumption 1-4}, we get the time-derivative of $\dot{\mathcal{V}}$:
\begin{equation}
\begin{aligned}
\label{dtLyap}
\dot{\mathcal{V}}&=\mathbf{z}^{\top}(  A_{\text{cl}}^\top P + PA_{\text{cl}}  )\mathbf{z}+2\mathbf{z}^{\top}PB(\hat{ \mathbf{R}}( \mathbf{x}_{k},\mathbf{u}_{k} ) -\hat{ \mathbf{R}}( \mathbf{x}_{k},\mathbf{u}_{k-1} )+{\epsilon}(\mathbf{x}_{k},\mathbf{u}_{k}))\\
&\leq \lambda_{\text{max}}( A_{\text{cl}}^\top P + PA_{\text{cl}} )\|\mathbf{z} \|^{2}+2\lambda_{\text{max}}(P)\sigma(B)\|\mathbf{z} \|( L_{R}\|\mathbf{u}_{k}-\mathbf{u}_{k-1}\| +\epsilon_{m} ),
\end{aligned}
\end{equation}

\noindent Let $\lambda=-\lambda_{\text{max}}(PA_{\text{cl}}+ A_{\text{cl}}^\top P )$, $c_{1}=\lambda_{\text{min}}(P)$, $c_{2}=\lambda_{\text{max}}(P)$, $c_{3}=2\lambda_{\text{max}}(P)\sigma(B)$ denote the constants and consider the inequality $c_{1}\|\mathbf{z} \|^{2} \leq \mathcal{V} \leq c_{2}\|\mathbf{z} \|^{2}  $. Then (\ref{dtLyap}) boils down to
\begin{equation}
\label{LyapIneq}
\dot{\mathcal{V}}\leq -\frac{1}{c_{2}}(\lambda - \frac{c_{2}c_{3}}{c_{1}}\rho L_{R})\mathcal{V}+\frac{c_{3}}{\sqrt{c_{1}}}\epsilon_{m} \sqrt{\mathcal{V}}.
\end{equation}

\noindent Here, we define $\mathcal{W}=\sqrt{V},\dot{\mathcal{W}}=\dot{\mathcal{V}}/(2\sqrt{\mathcal{V}})$ to apply 
the Comparison Lemma \citep{khalilbook} and obtain the convergence $\lim_{t\to\infty} \|\Lambda\mathbf{z} \|$ with $\Lambda$ being the decomposition of $P=\Lambda^\top \Lambda$:
\begin{equation}
\begin{aligned}
\|\Lambda \mathbf{z}(t) \| &\leq \|\Lambda \mathbf{z}(t_{0}) \| \exp{(-\frac{1}{2c_{2}}(\lambda- \frac{c_{2}c_{3}}{c_{1}}\rho L_{R})(t-t_{0}))}+\frac{c_{2}c_{3}\sqrt{c_{1}}}{c_{1}\lambda-c_{2}c_{3}\rho L_{R}}\epsilon_{m}.
\end{aligned}
\end{equation}
\noindent This gives us $\|\mathbf{z}(t)\| \leq r_\mathbf{z}$ with $r_\mathbf{z}=\sigma(\Lambda^{-1})(\|\Lambda \mathbf{z}(t_{0}) \| + \frac{c_{2}c_{3}\sqrt{c_{1}}}{c_{1}\lambda-c_{2}c_{3}\rho L_{R}}\epsilon_{m})$
and from (\ref{FRIDAY}),

\begin{equation}
\|\mathbf{u}(t)\|\leq r_\mathbf{u},\quad r_\mathbf{u}=\sigma(K)r_\mathbf{z}+u_{r_m}+ \epsilon_{m}+ R_{m}.
\end{equation}

\noindent As long as $r_\mathbf{z}\leq r_\mathcal{X}$ and $r_\mathbf{u}\leq r_\mathcal{U}$, $\mathbf{x}, \mathbf{u}$ lies exclusively inside $\mathcal{X}, \mathcal{U}$, 
yeilding $\|\mathbf{z}(t)\| \to \bar{B}_{r}(\mathbf{0},r=\sigma(\Lambda^{-1})\frac{c_{2}c_{3}\sqrt{c_{1}}}{c_{1}\lambda-c_{2}c_{3}\rho L_{R}}\epsilon_{m} )$.
This result also implies a theoretical trade-off: smaller $L_R$ conveges $\mathbf{z}$ faster but leaves a larger offset.$\quad\Box$


\section{Experiments}
\indent We evaluate FRIDAY performance in trajectory tracking simulation. The experimental setup is where FRIDAY only knows a nominal model and learns the map of uncertain disturbances while controlling the system. In this experiment, a simple mass system is used as the nominal model, and a nonlinear term is added to it as \textit{truth-model}.
First, we show FRIDAY improves the tracking performance compared to a Baseline controller. Next, we demonstrate that SN guarantees the closed-loop stability by comparing the estimation error of FRIDAY without SN. Lastly, we present an intuition about how SN-DNNs are suitable for FRIDAY compared to another popular data-driven estimator, GPs.
All our experiments are performed on Google Collaboratory with 12 GB of RAM and a 2.20GHz Intel Xeon processor. The code is available on: \texttt{\url{https://github.com/SpaceTAKA/FRIDAY_CarSimu}}
\subsection{Defining Nominal Model}
\indent For simplicity, we set the following mass system as a nominal model, which can be seen as a longitudinal car \citep{khalilbook}
\begin{equation}
m\ddot{\text{p}}=\text{u},
\end{equation}
\noindent where $\ddot{\text{p}}, \text{u}, m$ are the longitudinal acceleration, the driving force and the mass of the vehicle, weighing 1.5 kg respectively. We write it down to the LTI system $\dot{\mathbf{x}}= A\mathbf{x}+B\text{u}$, with $\mathbf{x}=[\text{p},\dot{\text{p}}]^{\top}$, $A=\begin{bmatrix} 0&1 \\ 0&0 \end{bmatrix} $, $B=\begin{bmatrix} 0 \\ \frac{1}{m} \end{bmatrix} $. 
\subsection{Defining Truth-Models}
\indent Based on the nominal model, we define true nonlinear models named \textit{truth-models}, which the controllers do not know.\\
\indent \textit{Param-truth:} This model represents a situation where all parameters of the vehicle are indeed time-variant. We consider time-variant load mass $m(t)$ with $a_{load}=9$ and time-variant force coefficient $\lambda(t)$ with $T$ being simulation period \citep{khalilbook}:
\begin{equation}
\begin{aligned}
\label{ParamTruth}
m(t)\ddot{\text{p}}&=\lambda(t)\text{u},\quad
\left\{ 
  \begin{alignedat}{2}   
    m(t)&=m+a_{load} m(1-\text{e}^{-\frac{t}{T}}) \\   
    \lambda(t)&=\text{e}^{-\frac{t}{T}}
  \end{alignedat} 
\right.
\end{aligned}
\end{equation}


\indent \textit{Multi-truth:} This model represents a situation where the additive nonlinearity has multiplicative functions of control input $u$. We consider the additive uncertainty $\dot{\text{p}}^{2}\text{u} + \text{p}^{2} +  \dot{\text{p}}|\text{u}|$ \citep{Multi}:
\begin{equation}
\label{MultiTruth}
m\ddot{\text{p}}=(1 + \dot{\text{p}}^{2}  )\text{u} + \text{p}^{2} +  \dot{\text{p}}|\text{u}|.
\end{equation}
\indent \textit{Enviro-truth:} This model represents a situation where the vehicle is exposed to complex nonlinear environmental force. We consider air drag force $f_{air}$ with $c_{air}=0.6$ and rolling resitance $f_{roll}$ on icy road with $\mu_{icy}=0.6, a_{roll}=0.4,r_{1}=0.2,r_{2}=0.1$. To increase the nonlinearity, Duffing spring force is added with $k_{1}=0.5, k_{2}=0.3$ \citep{slip, roll}:
\begin{equation}
\begin{aligned}
\label{EnviroTruth}
m\ddot{\text{p}}&=\mu_{icy} \text{u} - f_{air}-f_{roll}-f_{duff},\quad
\left\{ 
  \begin{alignedat}{3}   
    f_{air}&=c_{air}\dot{\text{p}}^{2}\sin{\dot{\text{p}}},\\
    f_{roll}&= -\text{sign}(\dot{\text{p}})mg(r_{1}(1-\text{e}^{-a_{roll}|\dot{\text{p}} | })+r_{2}|\dot{\text{p}}|),\\
    f_{duff}&= k_{1}\text{p}+k_{2}\text{p}^{3}
  \end{alignedat} 
\right.
\end{aligned}
\end{equation}

\subsection{Implementation of FRIDAY}
\indent In this experiment, our DNN $\hat{ \mathbf{R}}( \mathbf{x},\text{u} )$ consists of four fully-connected hidden layers with the input and output dimensions 3 to 1. The number of neurons in each layer is 50. 
We spectrally normalize the Lipschitz constant $L_{R} =1$. The teacher data $\mathbf{y}$ for the training is the observed residual dynamics value using the relation $ \mathbf{R}( \mathbf{x},\text{u} )=[0 ,\ddot{\text{p}}-\frac{1}{m}\text{u}] $.
The weights are $Q = \text{diag}(20,5), R = 1$ and the control rate is $20$Hz.
\subsection{Baseline Controller}
\indent We compare FRIDAY to an adaptive FBL-LQR controller with guaranteed stability \citep{MRAC}. Instead of the DNN, this Baseline Controller approximates the uncertainty using weighted basis functions $\hat{W}(t)\boldsymbol{\sigma}(\mathbf{x},\mathbf{u})$, with update rate $\dot{\hat{W}}(t)=\gamma\boldsymbol{\sigma}\mathbf{e}PB$. Note that the learning rate $\gamma=0.03$ and $\mathbf{e}$ is the error between the state and reference model. 
We also compare to a simple LQR controller which can be seen as FRIDAY without the SN-DNN. \color{black}

\begin{figure}
  \centering
  \includegraphics[width=\linewidth]{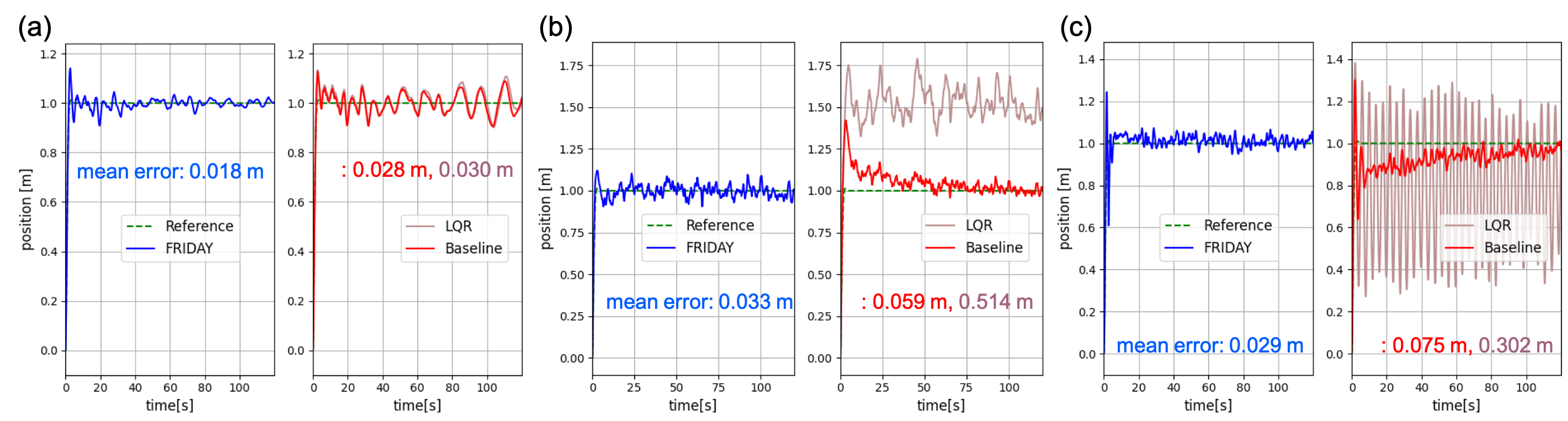}
  \caption{Setpoint regulation trajectory and the mean tracking error of 10 trajectories using FRIDAY (Left, blue), the Baseline controller (Right, red) and the LQR (Right, pink) in (a) \textit{Param-truth}, (b) \textit{Multi-truth}, and (c) \textit{Enviro-truth}.}
  \label{setpoint}
\end{figure}

\subsection{Tracking Performance}
\begin{figure}
  \centering
  \includegraphics[width=\linewidth]{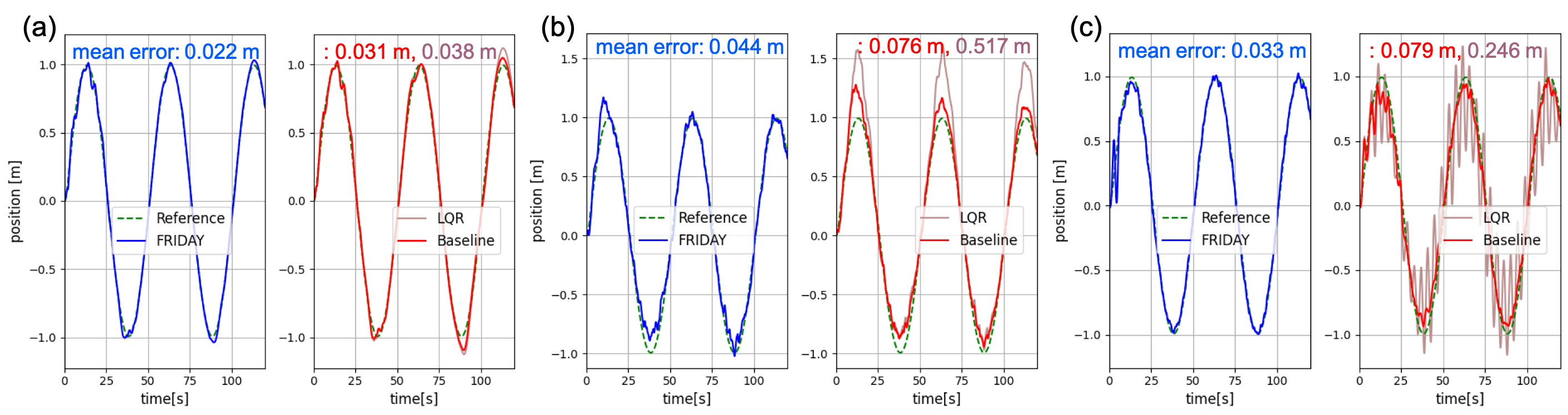}
  \caption{Sine wave tracking trajectory and the mean tracking error of 10 trajectories using FRIDAY (Left, blue), the Baseline controller (Right, red) and the LQR (Right, pink) in (a) \textit{Param-truth}, (b) \textit{Multi-truth}, and (c) \textit{Enviro-truth}.}
  \label{TrajeTest}
\end{figure}

\indent First, we conduct a setpoint regulation test. The target setpoint is $\mathbf{x}_r=[1,0]^{\top}$ with the reference control input $\text{u}_r=0$. 
From Fig. \ref{setpoint}, we can conclude that FRIDAY quickly and precisely reaches the target setpoint while the Baseline Controller slowly converges and the LQR shows a large offset (\ref{setpoint}-b) or highly oscillates (\ref{setpoint}-c) due to the unknown nonlinearity. FRIDAY achieves almost double the tracking accuracy of the Baseline Controller and ten times that of the LQR.\color{black} \\
For more practical use, we give a sine wave reference trajectory $\mathbf{x}_r=[\sin\omega t, \omega\cos\omega t]^{\top}$, $\text{u}_r=-m\omega^{2}\sin\omega t$ where the wave period $\omega = 2\pi/50$ [rad/s].
In Fig. \ref{TrajeTest}, similar to the setpoint regulation test, our controller outperforms the Baseline Controller and the LQR in tracking accuracy and the speed of convergence. FRIDAY smoothly fits the curve even when the Baseline and the LQR struggle to follow the path.\color{black}

\subsection{DNN Prediction Performance}
\begin{figure}
  \centering
  \includegraphics[width=0.7\linewidth]{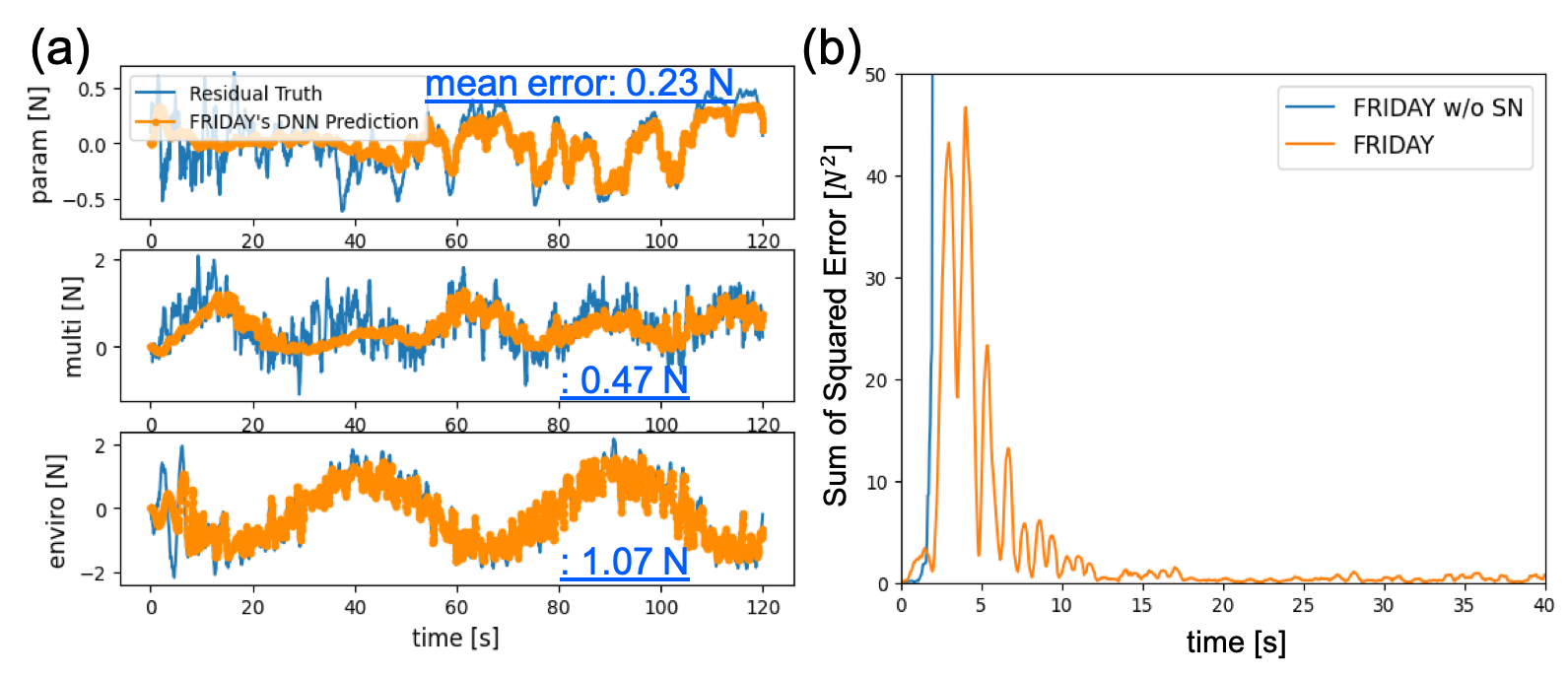}
  \caption{(a) Real-time estimated residual dynamics $\hat{ \mathbf{R}}$ compared to the true dynamics $\mathbf{R}$ with mean estimation error in the sine wave tracking case (see Fig. \ref{TrajeTest}). (b) Learning Loss of FRIDAY and FRIDAY without SN in the sine wave/\textit{Enviro-truth} case with $r_{1}=0.4, a_{roll}=0.8$. }
  \label{PrePerf}
\end{figure}

\indent The comparison of the true dynamics $\mathbf{R}$ and the predicted dynamics $\hat{ \mathbf{R}}$ is shown in Fig. \ref{PrePerf} (a). We observe the FRIDAY learns the mapping more and more precisely as time passes by. 
Though in some curves, the DNN prediction does not fit, this is because the curvatures exceed the bounded Lipschitz constant. 
Without SN, we can see FRIDAY's learning error overflows in Fig. \ref{PrePerf} (b), which empirically implies the necessity of SN to stabilize the closed-loop system.

\begin{table}[hbtp]
  \caption{Comparison of learned models }
  \label{estimators}
  \centering
  \scalebox{0.8}[0.8]{
  \begin{tabular}{lcr}
    \hline
    Model  & Wall time[s]  &  Mean estimation error[N]  \\
    \hline \hline
    SN-DNN  & 1.9  & 1.14 \\
    DNN  & 1.8  & 1.82 \\
    GP  & 104.0   & 1.10\\

    \hline
  \end{tabular}
  }
\end{table}

\subsection{Estimator Comparison}
\indent To provide an intuition about how SN-DNNs are suitable for FRIDAY's real-time estimation, we compare an SN-DNN with another popular data-driven estimator GPs, in the light of the training runtime and the mean estimation error. We use a GP model from Python library GPy with kernel Matern52. 
The training data is collected using the LQR controller following random setpoints (between 1.0 m and -1.0 m). 
After the training, we implement those learned models to our controller in the sine wave/\textit{Enviro-truth} case.\\
\indent From Table \ref{estimators}, we can find two main preferences of SN-DNNs for FRIDAY. (a) the SN-DNN learns the residual dynamics much faster than the GP and exhibits relatively good estimation. (b) the SN-DNN generalizes better than the one without SN. These preferences are supported by the fact that real-time learning SN-DNN in Fig. \ref{PrePerf} (a) shows a smaller mean estimation error than that of the learned SN-DNN in Table \ref{estimators}.

\section{Conclusion}
\indent In this letter, we present FRIDAY, composed of an LQR controller and a real-time stable learning DNN feedforward cancellation of unknown nonlinearity. Our framework shows two main benefits: 
(1) just by spectrally normalizing the weights, FRIDAY can learn fast with SGD and compensate for uncertain disturbances while controlling the system,  and (2) the stability of FRIDAY is strictly theoretically guaranteed.
Future works will include further application and generalization of FRIDAY.

\acks{I thank Guanya Shi, Xichen Shi, people on Mathematics Stack Exchange, and my family.}

\bibliography{l4dc2024FRIDAY}
\end{document}